\documentclass[twocolumn,superscriptaddress,nofootinbib,aps,floatfix,amsfonts,preprintnumbers,amsmath,amssymb,groupedaddress]{revtex4} 
\usepackage{graphicx, amsthm, multirow} 
\usepackage{url}
\usepackage{color}
\usepackage{dsfont}
\usepackage{stackrel}
\usepackage{bbm}
\usepackage{fancybox}
\usepackage{mathtools}
\usepackage{mdframed}
\usepackage{physics}
\usepackage{float}
\usepackage{bbold}
\usepackage{lpic}
\usepackage{array}
\usepackage{hyperref}

\definecolor{light-gray}{gray}{0.65}

\newcommand{\mb}{MiniBooNE }
\newcolumntype{L}[1]{>{\raggedright\let\newline\\\arraybackslash\hspace{0pt}}m{#1}}

\def\Utb{\ensuremath{|U_{\tau 2}|}}

\def\Uua{\ensuremath{|U_{\mu 1}|}}
\def\Uub{\ensuremath{|U_{\mu 2}|}}
\def\Uuc{\ensuremath{|U_{\mu 3}|}}

\def\Ueb{\ensuremath{|U_{e 2}|}}
\def\Uec{\ensuremath{|U_{e 3}|}}
\def\enorm{\ensuremath{|U_{e 1}|^2+|U_{e 2}|^2+|U_{e 3}|^2}}
\def\tnorm{\ensuremath{|U_{\tau 1}|^2+|U_{\tau 2}|^2+|U_{\tau 3}|^2}}
\def\unorm{\ensuremath{|U_{\mu 1}|^2+|U_{\mu 2}|^2+|U_{ \mu 3}|^2}}

\begin{document}

\title{Unitarity and the three flavour neutrino mixing matrix.}

\author{Stephen Parke}
\affiliation{Theoretical Physics Department, Fermi National Accelerator Laboratory, P.O.Box 500, Batavia, IL 60510, USA}
\author{Mark Ross-Lonergan}
\affiliation{IPPP, Department of Physics, Durham University, Durham DH1 3LE, UK}

\begin{abstract} 
Unitarity is a fundamental property of any theory required to ensure we work in a theoretically consistent framework. In comparison with the quark sector, experimental tests of unitarity for the 3x3 neutrino mixing matrix are considerably weaker. It must be remembered that the vast majority of our information on the neutrino mixing angles originates from $\overline{\nu}_e$ and $\nu_\mu$ disappearance experiments, with the assumption of unitarity being invoked to constrain the remaining elements. New physics can invalidate this assumption for the 3x3 subset and thus modify our precision measurements. We perform a reanalysis to see how global knowledge is altered when one refits oscillation results without assuming unitarity, and present $3 \sigma$ ranges for allowed $U_\text{PMNS}$ elements consistent with all observed phenomena. We calculate the bounds on the closure of the six neutrino unitarity triangles, with the closure of the $\nu_e \nu_\mu$ triangle being constrained to be $\leq$ 0.03, while the remaining triangles are significantly less constrained to be $\leq$ 0.1 - 0.2.  Similarly for the row and column normalization, we find their deviation from unity is constrained to be $\leq$ 0.2 - 0.4, for four out of six such normalisations, while for the $\nu_\mu$ and $\nu_e$ row normalisation the deviations are constrained to be $\leq$ 0.07, all at the  $3\sigma$ CL. We emphasise that there is significant room for new low energy physics, especially in the $\nu_\tau$ sector which very few current experiments constrain directly.

\end{abstract} 


\maketitle

\begin{table*}[]
{\renewcommand{\arraystretch}{1.4}
    \begin{tabular}{ | L{2.6cm} | L{5.5cm} | l | l |} 
    \hline
{\bf Experiment }&{\bf Measured quantity with unitarity }&{\bf Without unitarity}&{\bf Normalisation} \\ \hline \hline
Reactor SBL  ($\overline{\nu}_e\rightarrow \overline{\nu}_e$)  & $4|U_{e3}|^2\left(1-|U_{e3}|^2\right) = \sin^2 2\theta_{13}$ & $4|U_{e3}|^2\left( U_{e1}|^2 +|U_{e2}|^2\right)$ & $ \left( \enorm \right)^2$  \\ \hline  \hline 
Reactor LBL ($\overline{\nu}_e\rightarrow \overline{\nu}_e$)  & $4|U_{e1}|^2 |U_{e2}|^2  =  \sin^2 2 \theta_{12} \cos^4 \theta_{13} $ &  $4|U_{e1}|^2 |U_{e2}|^2$ & $ \left( \enorm \right)^2$ \\ \hline \hline
SNO ($\phi_{CC}/\phi_{NC}$ Ratio) & $|U_{e2}|^2  =  \cos^2 \theta_{13} \sin^2 \theta_{12} $ & $|U_{e2}|^2 $ & $ \Ueb^2+\Uub^2+\Utb^2 $\\ \hline \hline
SK/T2K/MINOS ($\nu_\mu \rightarrow \nu_\mu$)  & $4|U_{\mu 3}|^2\left(1-|U_{\mu 3}|^2\right)  = 4 \cos^2\theta_{13} \sin^2\theta_{23}\left(1-\cos^2\theta_{13}\sin^2\theta_{23}\right)$  & $4|U_{\mu3}|^2 \left( U_{\mu1}|^2 +|U_{\mu2}|^2\right)$ & $ \left( \unorm \right)^2$\\ \hline \hline
T2K/MINOS ($\nu_\mu \rightarrow \nu_e$)  & $4|U_{e 3}|^2|U_{\mu 3}|^2 =  \sin^2 2 \theta_{13} \sin^2 \theta_{23}$ & $-4 \Re \lbrace U_{e 3}^* U_{\mu 3}  \left( U_{e1}^* U_{\mu 1}+U_{e2}^* U_{\mu 2}  \right) \rbrace $ & $ |U_{e 1} U_{\mu 1}^* +U_{e 2} U_{\mu 2}^*+U_{e 3} U_{\mu 3}^*|^2 $\\ \hline \hline
SK/OPERA ($\nu_\mu \rightarrow \nu_\tau$) & $4|U_{\mu 3}|^2|U_{\tau 3}|^2  =  \sin^2 2 \theta_{23} \cos^4 \theta_{13}$ & $-4 \Re \lbrace U_{\tau 3}^* U_{\mu 3}  \left( U_{\tau 1}^* U_{\mu 1}+U_{\tau 2}^* U_{\mu 2}  \right) \rbrace $ & $ |U_{\mu 1} U_{\tau 1}^* +U_{\mu 2} U_{\tau 2}^*+U_{\mu 3} U_{\tau 3}^*|^2 $\\ \hline 
    \end{tabular}}
\caption{Example experiments and the leading order functions of $U_\text{PMNS}$ matrix elements they measure, in both the unitary and non-unitary case. The third column shows the normalisation that can be bound if the experimental measurements of the fluxes and backgrounds are known to a high enough degree.}
\label{fig:pmns} 
\end{table*}
With the knowledge of $\sin^2 2 \theta_{13}$ now almost at the 5\% level, and interplay between the long baseline accelerator $\nu_\mu \rightarrow \nu_e$ appearance data \cite{Abe:2015awa,Adamson:2013ue} and short baseline reactor $\overline{\nu}_e \rightarrow \overline{\nu}_e$ disappearance \cite{An:2012eh,Ahn:2012nd,Abe:2012tg} data, combined with prior knowledge of $\theta_{23}$ from $\nu_\mu \rightarrow \nu_\mu$ disappearance data \cite{Wendell:2010md,Sousa:2015bxa,Abe:2014ugx}, suggesting tentative global hints at $\delta_{CP} \approx 3 \pi /2$, there is much merit to statements that we are now in the precision measurement era of neutrino physics. Our knowledge of the distinct Pontecorvo-Maki-Nakagawa-Sakata (PMNS) neutrino mixing matrix elements comes from the plethora of successful experiments that have run since the first strong evidence for neutrino oscillations, interpreted as $\nu_\mu \rightarrow \nu_\tau$ oscillations, was discovered by Super-Kamiokande in 1998 \cite{Fukuda:1998mi}. However, one must always remember that our knowledge of the matrix elements is predominately in the $\nu_e$ and $\nu_\mu$ sectors, and comes primarily from high statistics $\overline{\nu}_e$ disappearance and $\nu_\mu$ disappearance experiments, with the concept of unitarity being invoked to disseminate this information onto the remaining elements.  With more statistics, the long baseline $\nu_\mu \rightarrow \nu_e$ and $\overline{\nu}_\mu \rightarrow \overline{\nu}_e$ appearance experiments such as T2K \cite{Abe:2013hdq} and NO$\nu$A \cite{Ayres:2004js} will aid in $\nu_\mu$ sector precision measurements.

Unitarity of a mixing matrix is a necessary condition for a theoretically consistent description of the underlying physics, as non-unitarity directly corresponds to a violation of probability in the calculated amplitudes. In the neutrino sector unitarity can be directly verified by precise measurement of each of the mixing elements to confirm the unitarity condition:  $U^\dagger U = \mathds{1} = U U^\dagger$. In this there are 12 dependant conditions, six of which we will refer to as normalisations (sum of the squares of each row or column, e.g the $\nu_e$ normalisation $\enorm=1$) and six conditions that measure the degree with which each unitarity triangle closes (e.g the $\nu_e \nu_\mu$ triangle:  $ U_{e 1} U_{\mu 1}^* +U_{e 2} U_{\mu 2}^*+U_{e 3} U_{\mu 3}^* =0 $). Currently, from direct measurements of the individual elements only, the $\nu_e$ normalisation is the sole condition that can be reasonably constrained without any further assumptions as to the origin of the non-unitarity \cite{Qian:2013ora}.

In the quark sector, the analogous situation involving the Cabibbo-Kobayashi-Maskawa (CKM) matrix has been subject to intense verification as many distinct experiments have access to probes of all of the $V_\text{CKM}$ elements individually. Current data shows that the assumption of unitarity for the 3x3 CKM matrix is valid in the quark sector to a high precision, with the strongest normalisation constraint being $|V_{ud}|^2+|V_{us}|^2+|V_{ub}|^2 = 0.9999 \pm 0.0006$ and the weakest still being significant at $|V_{ub}|^2+|V_{cb}|^2+|V_{tb}|^2 = 1.044 \pm 0.06$ \cite{Agashe:2014kda}. Unlike the quark sector, however, experimental tests of unitarity are considerably weaker in the 3x3 $U_\text{PMNS}$ neutrino mixing matrix. It remains an initial theoretical assumption inherent in many analyses \cite{Gonzalez-Garcia:2014bfa,Fogli:2012ua,Forero:2014bxa}, but is the basis for the validity of the $3\nu$ paradigm. 

This non-unitarity can arise naturally in a large variety of theories. A generic feature of many Beyond the Standard Model scenarios is the inclusion of one or more new massive fermionic singlets, uncharged under the Standard Model (SM) gauge group, $SU(3)_C \times SU(2)_L \times U(1)_Y$. If these new sterile states mix with the SM neutrinos then the true mixing matrix is enlarged from the 3x3 $U_\text{PMNS}$ matrix to a $n$x$n$ matrix, 
\begin{figure}[h!]
\vspace*{-0.3cm}
        \centering
  \includegraphics[width=0.45\textwidth]{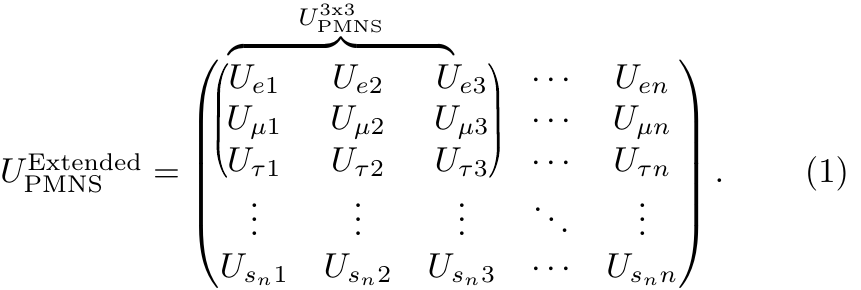}
        \label{fig:visual}
\end{figure} 

\setcounter{equation}{1}
These so-called sterile neutrinos have been a major discussion point for both the theoretical and experimental communities for decades. If they have masses at or near the GUT scale, then the see-saw mechanism can provide sufficiently small masses to the SM neutrinos \cite{Minkowski:1977sc}, but a priori these new states can sit at practically any mass as there is no known symmetry to dictate a scale. Although this extended $n$x$n$ mixing matrix, should nature choose it, will indeed be unitary to preserve probability, the same is not true for any given $m$x$m$ subset, with $m<n$. This is the canonical model of how new physics, introduced at any scale, breaks observed unitarity in the neutrino sector. \\ 

An extensive body of work in the literature exists on non-unitarity in the neutrino sector, most of which has been analysed with the rigorous model-independent approach of the Minimal Unitarity Violation (MUV) scheme \cite{Antusch:2006vwa}. In this approach the new physics enters high above the energies involved in oscillation experiments, and the three neutrino Standard Model becomes a low-energy effective theory in which the unitarity of the 3x3 mixing matrix is not assumed. It has been shown recently \cite{Antusch:2014woa} that the current status under this scheme is highly constrained by experiment, the weakest unitarity constraint is $|U_{\tau 1}|^2 +|U_{\tau 2}|^2+|U_{\tau 3}|^2 = 0.9947 \rightarrow 1.0$ at the 90 \% CL, producing practically immeasurable deviations to the mixing angles with current experimental uncertainty. Some of the most stringent bounds in the MUV scheme come from rare lepton decays such as $\mu \rightarrow e \gamma$. This is due to the fact that without the unitarity condition in the 3x3 mixing matrix, the exact (in the massless $\nu$ limit) cancellation provided by the Glashow-Iliopoulos-Maiani mechanism \cite{PhysRevD.2.1285} in the SM no longer holds. In conjunction with the fact each pair of active neutrinos and charged leptons, e.g. $(\nu_e, e_L)$, make up $SU(2)_L$ electroweak doublets, this results in the off diagonal terms of ${U_\text{PMNS}^\dagger U_\text{PMNS}}$ being hugely constrained by the current experimental limits on the branching ratio $BR(\mu \rightarrow e \gamma)$. The strongest limits currently arise from the MEG experiment, $BR(\mu \rightarrow e \gamma) <  5.7 \times 10^{-13}$ at the 90\% CL \cite{Adam:2013mnn} which translates to a bound on $|U_{e1}^*U_{\mu 1}+U_{e2}^*U_{\mu 2}+U_{e3}^*U_{\mu 3} |<10^{-5}$ \cite{Antusch:2014woa} at the 90\% CL.

If, however, the new physics that provides the non-unitarity enters at a much lower energy scale, as several current experimental hints suggest with anomalous results from LSND \cite{Aguilar:2001ty}, \mb \cite{Aguilar-Arevalo:2013pmq}, the Gallium anomaly \cite{Hampel1998114,Abdurashitov:1998ne} and the Reactor anomaly \cite{Mention:2011rk}, then many of the most constraining experiments that bound the MUV scheme are not directly applicable. For physics entering at these lower scales, one must focus on direct measurements of the individual mixing elements. To this end neutrino oscillations are the most important experimental probe we have access to. The most convincing means of verification of unitarity in the neutrino sector would be analogous to the quark sector, via direct and independent measurement of all the $U_\text{PMNS}$ elements, to overconstrain the parameter space and confirm that the 12 unitarity constraints hold to within experimental precision. However, we do not currently have access to enough experiments, at distinct mass differences, in the $\nu_\mu$ and $\nu_\tau$ sectors to bound all of the elements to a sufficient degree to verify all 12 conditions.

The situation for progress in direct measurement is therefore not promising. Thus we must look for alternative ways to constrain the $U_\text{PMNS}$ elements. One can perform indirect searches of unitarity by searching for mixing elements outside those of the $3 \nu$ mixing regime. These class of searches do not measure the 3x3 mixing elements per say, but rather by looking for additional states one can constrain the violations they would induce in the 3x3 subset. One proceeds by noting all null results in oscillations at mass differences distinct to those of the $3 \nu$ paradigm. We do not wish to perform a global fit for new physics as this has been well covered in the literature \cite{Kopp:2013vaa,Conrad:2012qt}, instead we focus on what unresolved physics can do to our current precision, hence we do not include any positive signals such as LSND or the MiniBooNE anomaly. Such an sterile driven approach requires additional assumptions on the exact origin of the non-unitarity, thus losing some model-independence. However, as an extended $U_\text{PMNS}$ matrix encompasses many beyond the Standard Model scenarios, it is natural to include this in our analysis. To proceed one must then consider what scale the new physics enters at, however, as we do not focus on the origin of such non-unitarity we choose to marginalise over the new scale(s) assuming the possibility they enter in at an oscillating scale, with at least $|\Delta m^2| \geq 10^{-2} \text{ eV}^2$. Below this scale, states degenerate with SM neutrinos requires a much more detailed analysis.

A non-unitary mixing matrix can be parameterised as a 3x3 matrix hosting 9 complex non-unitary elements, 5 phases of which can be removed by rephasing the lepton fields, leaving 13 parameters: 9 real positive numbers and 4 phases. The oscillation probability for a neutrino (anti-neutrino) of initial flavour $\alpha$ and energy $E_\nu$ to transition to a neutrino  (anti-neutrino) of flavour $\beta$ after a distance $L$ with such a non-unitary mixing matrix is given by 
\begin{align}
P\left(\overset{\text{ \tiny (--) }}{\nu_\alpha} \rightarrow \overset{\text{ \tiny (--) }}{\nu_\beta}\right) &=  \left| \sum_{i=1} U_{\beta i}^* U_{\alpha i} \right|^2  \label{eq:prob}
\\
 & \hspace*{-15mm} - 4 \sum_{i<j} \Re(U_{\beta i}U_{\beta j}^* U_{\alpha i} U_{\alpha j }^*)\sin^2 \left(\Delta m_{ji}^2 \frac{L}{4 E_\nu}\right) \nonumber\\
 & \hspace*{-15mm} \overset{\text{ \tiny (---) }}{+} 2 \sum_{i<j} \Im(U_{\beta i}U_{\beta j}^* U_{\alpha i} U_{\alpha j }^*)\sin(\Delta m_{ji}^2 \frac{L}{2 E_\nu}),
\nonumber
\end{align}

where now, without assuming unitarity, the leading term is not a function of $\Delta  m^2 L/E_{\nu}$ and is also not necessarily equal to 1 or 0 in neutrino disappearance and appearance experiments respectively. The leading term directly probes three row normalisations in disappearance experiments and three row unitarity triangles in appearance. Table (\ref{fig:pmns}) shows some of the current most constraining experiments and the associated mixing matrix elements that they can directly measure. Although violations of unitarity such as these modify the oscillation amplitudes and total normalisation of the probability, they do not have any effect of the oscillation frequency which remains a function of the mass differences and $L/E_\nu$ only (ignoring higher order non-unitary matter effects). Thus, for simplicity of analysis the global best fit values for the mass squared differences are assumed ($\Delta m_{21}^2 = 7.6 \times 10^{-5} \text{eV}^2, |\Delta m_{31}^2| = 2.4 \times 10^{-3}  \text{eV}^2$) \cite{Agashe:2014kda}. For each observed oscillation one can then directly compare the measured amplitude with the non-unitary expression in equation (\ref{eq:prob}). It is this amplitude-matching that we use to undertake a global-fit and provides us the ranges for $U_\text{PMNS}$ that would successfully reproduce the measured oscillation amplitudes and normalisations. We focus on the physically motivated subclass of unitarity violations such that $ |U_{\alpha 1}|^2+|U_{\alpha 2}|^2+|U_{\alpha 3}|^2 \le1$, for $\alpha = e,\mu,\tau$, and $|U_{e i}|^2+|U_{\mu i}|^2+|U_{\tau i}|^2 \le1$ for $ i = 1,2,3$. 
Added assumptions on the exact origin of the non-unitarity in the 3x3 submatrix can lead to further correlations between the elements. In particular, if the non-unitarity does indeed come from an enlarged mixing matrix then one must invoke Cauchy-Schwartz inequalities along with the unitarity constraints of the true extended mixing matrix to place six geometric constraints on the mixing elements \cite{Antusch:2006vwa},
\begin{align}
\left| \sum_{i=1}^3 U_{\alpha i} {U_{\beta i}}^* \right|^2 &\leq \left(1-\sum_{i=1}^3 |U_{\alpha i}|^2\right)\left(1-\sum_{i=1}^3 |U_{\beta i}|^2\right),  \nonumber \\ 
& \qquad \text{for } \quad \alpha,\beta = (e,\mu,\tau), \quad \alpha \neq \beta, \nonumber  \\ 
\left| \sum_{\alpha=e}^\tau U_{\alpha i} {U_{\alpha j}}^* \right|^2 &\leq \left(1-\sum_{\alpha=e}^\tau |U_{\alpha i}|^2\right)\left(1-\sum_{\alpha=e}^\tau      |U_{\alpha j}|^2\right), \nonumber \\ 
&\qquad \text{for } \quad i,j = (1,2,3), \quad i \neq j .  
\label{eq:cond2}
\end{align}
These geometric constraints enable precision measurements in a single sector to be passed subsequently to all elements of the mixing matrix\footnote{These Cauchy-Schwartz inequalities are analogous to the statement that one can bound $\nu_\mu \rightarrow \nu_e$ appearance by the associated $\nu_\mu$ and $\nu_e$ neutrino disappearance limits, in 3+N sterile neutrino scenarios.}.

 \begin{figure*}
        \centering
  \includegraphics[width=0.93\textwidth]{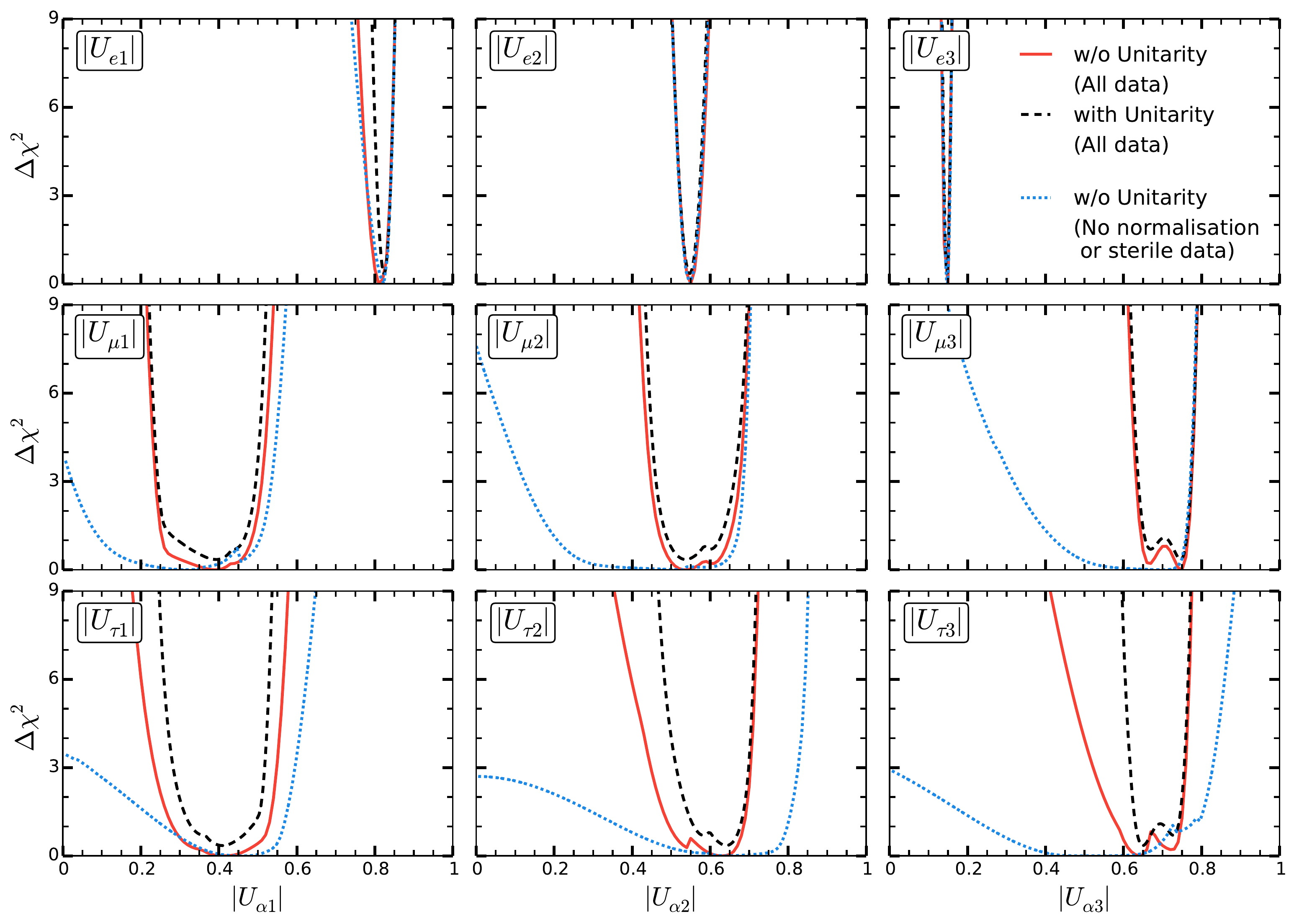}
        \caption{Marginalised 1-D $\Delta \chi^2$ for each of the magnitudes of the 3x3 neutrino mixing matrix elements, without (red solid) and with (black dashed) the assumption of unitarity, using all data. The x-axis is the magnitude of each individual matrix element, and the y-axis is the associated $\Delta \chi^2$ after marginalisation over all parameters other than the one in question. Also shown is the fit in which no normalisation or sterile search data is used  (blue dotted), to highlight their importance to the fit. The blue and red curves coincide closely with the unitary case in $\Ueb$ and $\Uec$.  This analysis was preformed for the normal hierarchy, the inverse hierarchy providing the same qualitative result.}
        \label{fig:visual}
\end{figure*}

To perform the analysis, for each experiment considered\footnote{The experimental data considered in this analysis is: Bugey \cite{Declais:1994su}, CCFR \cite{Stockdale:1985ce,Naples:1998va,Romosan:1996nh,McFarland:1995sr}, CDHS \cite{Dydak:1983zq}, CHORUS \cite{Eskut:2007rn}, CHOOZ \cite{Apollonio:1999ae}, Daya Bay \cite{An:2013uza,An:2014bik}, Double Chooz \cite{Abe:2012tg}, ICARUS \cite{Antonello:2013gut}, KARMEN \cite{Armbruster:2002mp}, KamLAND \cite{Abe:2008aa,Gando:2010aa}, MINOS \cite{Adamson:2013whj,Adamson:2013ue,Adamson:2010wi,Sousa:2015bxa}, NOMAD \cite{Astier:2003gs,Astier:2001yj} , NO$\nu$A \cite{novaresults}, NuTeV \cite{Avvakumov:2002jj}, OPERA \cite{Agafonova:2015jxn,Agafonova:2015neo}, RENO \cite{Ahn:2012nd}, SNO \cite{Aharmim:2011vm}, SciBooNE \cite{Cheng:2012yy}, Super-Kamiokande \cite{Wendell:2010md,Abe:2014gda,Toshito:2001dk,Koshio:2015dfa}, T2K \cite{Abe:2014ugx,Abe:2013hdq}.} we take the observed amplitude of the $\nu_\alpha \rightarrow \nu_\beta$ (or $\overline{\nu}_\alpha \rightarrow \overline{\nu}_\beta$) oscillation after normalisation, $A_{\alpha; \beta}$, alongside its published error $\sigma_{\alpha; \beta}$ and construct a $\chi^2 = (P^{NU}_{\alpha; \beta}-A_{\alpha; \beta})^2/\sigma_{\alpha:\beta}^2$ for the associated non-unitary amplitudes as taken from equation (\ref{eq:prob}), $P^{NU}_{\alpha; \beta}$, along with any necessary normalisation systematics as pull factors. For indirect short-baseline (SBL) and sterile searches, if an experiment publishes the resultant $\chi^2$ surface of their analyses in a 3+N format then this is used as a prior to bound any non-unitarity. Otherwise an appropriate prior is estimated by performing a 3+N fit to published data, with sterile masses allowing for the possibility of a sterile too light to oscillate yet or so heavy its oscillations have become averaged out and up to three steriles within the active oscillation region where spectral distortions are evident, to approximate a 3+N CP violating scenario. 

We minimize the constructed $\chi^2$ over all parameters, satisfying the Cauchy-Schwartz constraints given by equations (\ref{eq:cond2}), using an adaptive MCMC minimizer. The results of the analyses are shown in Fig. (\ref{fig:visual}), without unitarity (red solid line) and with the assumption of unitarity (black dashed line). The non-unitary analysis was performed under the strict assumption that any non-unitarity comes solely from an extended $U_\text{PMNS}$ and that no new interactions, such as an additional $U(1)'$ which can lead to strongly modified matter effects, are active at oscillation energies. 

Upon minimization the best fit points agree in both unitary and non-unitary fits. To compare how the precision varies we consider the frequentist $3 \sigma$ ranges of the one-dimensional $\Delta \chi^2$ projections without unitarity assumed (with unitarity), where we marginalise over all parameters except the one in question, we obtain 
\begin{align}
&|U|_{3 \sigma}^{\text{ \bf w/o Unitarity} \atop \text{(with Unitarity)}} =  \nonumber \\[1mm]
& \begin{pmatrix}
  0.76 \rightarrow 0.85   \hspace{0.5cm} &0.50 \rightarrow 0.60 \hspace{0.5cm} & 0.13 \rightarrow 0.16 \\[-1mm]
 {\scriptstyle ( 0.79 \rightarrow 0.85) }  \hspace{0.5cm} &  {\scriptstyle (0.50 \rightarrow 0.59) } \hspace{0.5cm} &  {\scriptstyle (0.14 \rightarrow 0.16) } \\[1mm]
   0.21 \rightarrow 0.54 \hspace{0.5cm} & 0.42 \rightarrow 0.70 \hspace{0.5cm} & 0.61 \rightarrow 0.79 \\[-1mm]
{\scriptstyle ( 0.22 \rightarrow 0.52) }  \hspace{0.5cm} &  {\scriptstyle (0.43 \rightarrow 0.70) } \hspace{0.5cm} &  {\scriptstyle (0.62 \rightarrow 0.79) } \\[1mm] 
  0.18 \rightarrow 0.58 \hspace{0.5cm}& 0.38 \rightarrow 0.72 \hspace{0.5cm} & 0.40 \rightarrow 0.78 \\[-1mm]
   {\scriptstyle ( 0.24 \rightarrow 0.54) }  \hspace{0.5cm} &  {\scriptstyle (0.47 \rightarrow 0.72) } \hspace{0.5cm} &  {\scriptstyle (0.60 \rightarrow 0.77) } \\[1mm]
 \end{pmatrix}.
 \end{align}
 The ranges for the individual elements, assuming unitarity (bracketed numbers in above expression),  are in good agreement with published results in contemporary global fits such as $\nu$-fit \cite{Gonzalez-Garcia:2014bfa}. \\

\begin{figure}
\centering
\includegraphics[width=0.47\textwidth]{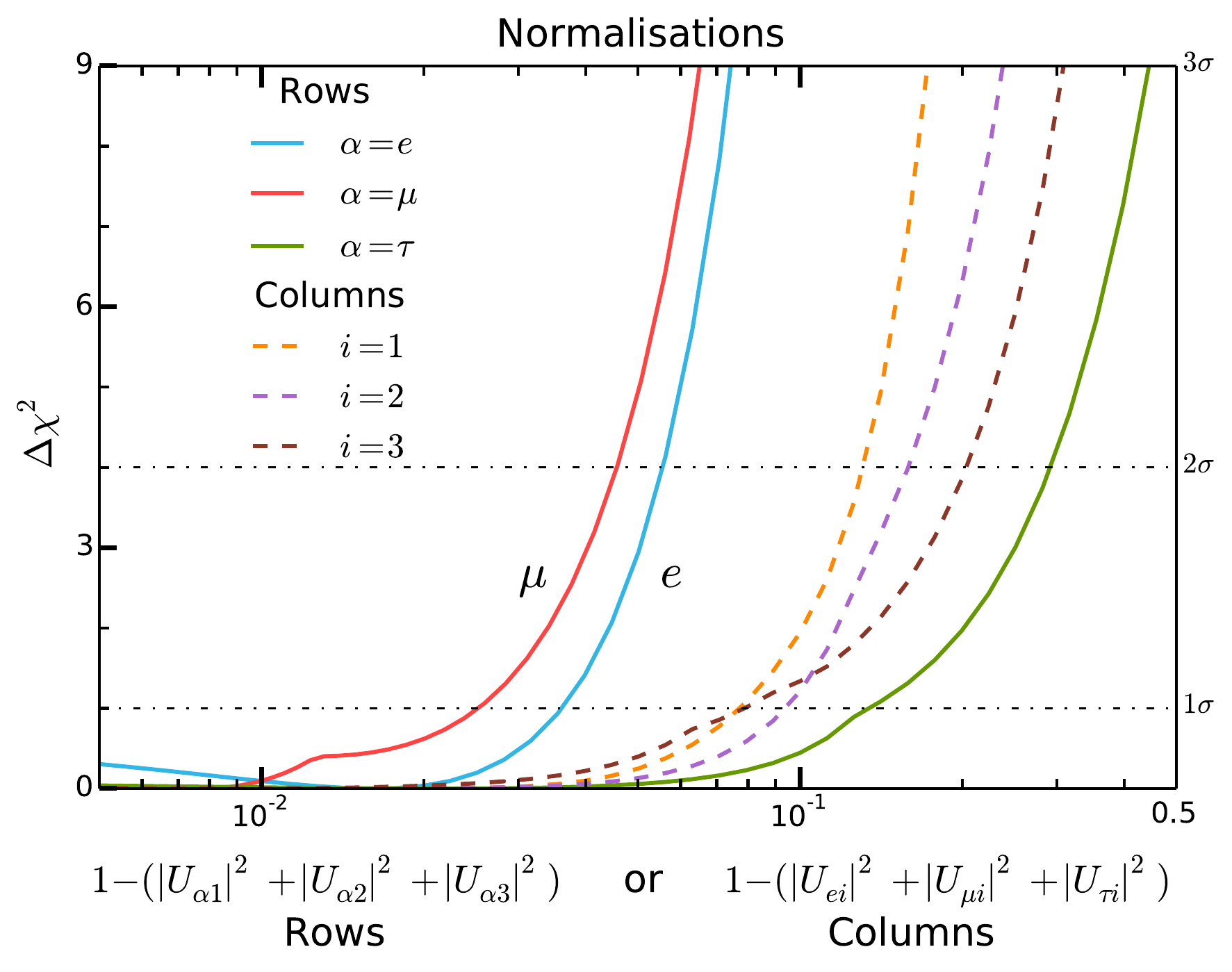}
        \caption{1-D $\Delta \chi^2$ for deviation of both $U_\text{PMNS}$ row (solid) and column (dashed) normalisations, fitted with all spectral and normalisation data, when considering new physics that enters above $|\Delta m^2|\geq 10^{-2}$eV$^2$. }
        \label{fig:norm}
\end{figure}

Also in Fig. (\ref{fig:visual}),  for the purpose of emphasising how important normalisation and sterile searches are to the precision of the 3x3 $U_\text{PMNS}$ elements, without the assumption of unitarity, we performed an alternative, more conservative fit, in which no normalisation or 3+N sterile search data is included. This fit only highlights the stark difference between the quark and neutrino sectors. Whereas the quark sector can independently measure all $V_\text{CKM}$ elements, some within 0.1\% and all within 10\% \cite{Agashe:2014kda}, in the neutrino sector we only have access to precision knowledge of the $\nu_e$ row in a completely model-independent manner. This is as expected, as the $\nu_e$ sector has access to high statistics experiments at both the solar and atmospheric mass scale, as well as a third independent experiment in the solar flux measurements, due to the MSW effect. In the $\nu_\mu$ sector, $\Uuc^2$ is known to be non-zero from $\nu_\mu$ disappearance experiments but as $\Uua$ and $\Uub$ only appear in the degenerate combination $\Uua^2+\Uub^2$, they cannot be distinguished individually. This degeneracy is very weakly broken by the $\nu_\mu \rightarrow \nu_e$ appearance experiment T2K \cite{Abe:2015awa}, and will be improved upon taking of more data and with future high statistics NO$\nu$A \cite{Ayres:2004js} results. The addition of this normalisation and sterile data in the $3\nu$ unitarity case does not change anything in the fit. From here on we will discuss only the main results, as calculated including all normalisation and sterile search data.

The addition of this sterile search and normalisation data improves the situation significantly. If we define the shift in range of allowed values as the ratio of the difference in $3\sigma$ ranges without and with unitarity, to that derived with unitarity, the increase in parameter space for $|U_{e i}|, i=2,3$ and $|U_{\mu i}| , i=1,2,3$ are all $\leq$ 10\% (4\%, 8\%, 8\%, 7\% and  4\% respectively), with $|U_{e 1}|$ taking the majority of the discrepancy in the $\nu_e$ sector, with an increase of allowed range of 68\%, primarily due to the weaker bounds from KamLAND compared to the SBL reactors, and that $|U_{e1}|^2$ forms the bulk of $\enorm$. The entire $\nu_\tau$ sector, however, may contain substantial discrepancies from unitarity with shifts in allowed regions of 37\%, 46\% and 104\% respectively. We have little or no current mechanisms to directly measure any $\nu_\tau$ elements and we have not yet observed any oscillation amplitude peaks, even the recent $5 \sigma$ discovery of $\nu_\mu \rightarrow \nu_\tau$ at OPERA \cite{Agafonova:2015jxn} only sees the tail end of the 1$^\text{st}$ oscillation maximum and the observation of 5 events on a background of $0.25 \pm 0.05$ is not significant spectrally and can be equally be fit by a flat normalisation discrepancy. The precision we do have is driven by the fact large deviations here cause violations of unitarity too large in the $\nu_e$ and $\nu_\mu$ sectors, passed through by the geometric Cauchy-Schwartz constraints. \\

We must stress that even if the $3\sigma$ ranges of the $U_\text{PMNS}$ elements agree closely with the unitarity case, this does not equate to the neutrino mixing matrix being unitary. In the unitary case the correlations are much stronger and choosing an exact value for any one the mixing elements drastically reduces the uncertainty on the remaining elements. To better understand the level at which we know unitarity is conserved or not, we plot the resultant ranges for the normalisation in Fig (\ref{fig:norm}). We see that the $\nu_e$ and $\nu_\mu$ normalisation deviations from unity are relatively well constrained ($\leq$ 0.06 and 0.07 at $3 \sigma$ CL respectively), primarily by reactor fluxes and a combination of precision measurements of the rate and spectra of upward going muon-like events observed at Super-Kamiokande \citep{Abe:2014gda} and the multitude of long and short baseline accelerator $\nu_\mu \rightarrow \nu_\mu$ disappearance experiments. We note the $\nu_\mu$ normalisation deviation from unity is constrained slightly ($\approx 1\%$) better than the $\nu_e$ normalisation. This is due to the large theoretical error, 5\%, on total flux from reactors assumed \cite{Hayes:2013wra}. The remaining normalisation deviations from unity are all constrained to be $\lesssim $ 0.2 - 0.4 at $3 \sigma$ CL.\\

\begin{figure}
\centering
\includegraphics[width=0.47\textwidth]{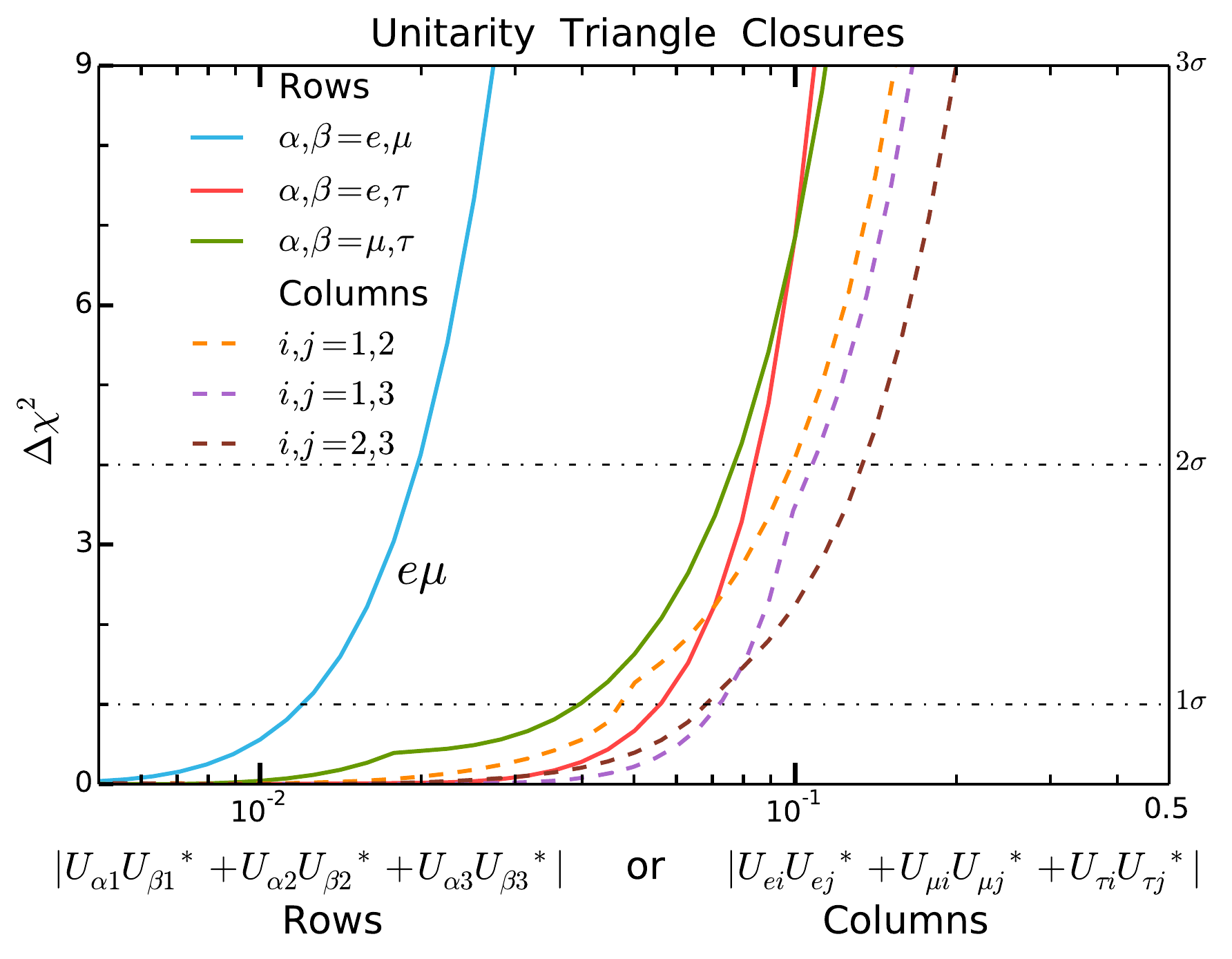}
        \caption{1-D $\Delta \chi^2$ for the absolute value of the closure of the three row (solid) and three column (dashed) unitarity triangles, fitted with all spectral and normalisation data, when considering new physics that enters above $|\Delta m^2|\geq 10^{-2}$ eV$^2$. There is one unique unitarity triangle, the $\nu_e\nu_\mu$ triangle, in that it does not contain any $\nu_\tau$ elements and hence is constrained to be unitary at a level half an order of magnitude better than the others. By comparison to Fig. \ref{fig:norm} one can clearly see the Cauchy-Schwartz constraints are satisfied.}
        \label{fig:triangle}
\end{figure}

For the case of the six neutrino unitarity triangles, we present the allowed ranges for their closures in Fig. (\ref{fig:triangle}). For the three row triangles the bounds originate from a combination of the corresponding geometric constraints along with appearance data in the respective channel. The column triangles, however, are bound by the geometric constraints only, and as the column normalisations are proportionally less known, so too are the column unitarity triangles. Only one triangle does not contain a $\nu_\tau$ element, the $\nu_e \nu_\mu$ triangle, and hence it is the only triangle in which it is excluded to be open by more than 0.03 at the $3\sigma$ CL, compared to between 0.1 - 0.2 at the $3\sigma$ CL for the remaining triangles. This hierarchical situation will not improve unless precise measurements can be made in the $\nu_\tau$ sector.  \\

If one wishes to proceed with measurements of unitarity, without the assumption of an extended $U_\text{PMNS}$ matrix and its subsequent Cauchy-Schwartz bounds, then prospects for improvement are essentially limited to measuring the $\nu_e$ normalisation. Improvement of all $\nu_e$ elements is possible, especially if the new generation reactor experiments, JUNO \cite{Li:2013zyd} and RENO50 \cite{reno50}, proceed as planned. See discussion by X. Qian et al. \cite{Qian:2013ora} for a detailed discussion of the possible improvements. Significant improvement in the $\nu_\mu$ sector would require the measurement of $\nu_\mu$ disappearance at the solar mass scale, well beyond what is currently technologically feasible.

Improvements in the indirect 3+N sterile measurements are much more promising, the Fermilab Short Baseline Neutrino (SBN) \cite{Antonello:2015lea} program consisting of the SBND, MicroBooNE and ICARUS experiments on the Booster beam, will be capable of probing a wide range of parameter space for 3+N models, increasing both the appearance and disappearance bounds. Subsequently, the long baseline program DUNE \cite{Adams:2013qkq} will also be able to significantly extend the constrained region of $\nu_\mu \rightarrow \nu_e$ appearance to lower mass differences, leading to increased constraints on the $\nu_e\nu_\mu$ unitarity triangle in this regime. An understanding of the neutrino flux and cross sectional uncertainties are crucial for unitarity measurements. Possible future experiments such as a fully fledged Neutrino Factory \cite{Choubey:2011zzq} or the nuStorm facility \cite{Adey:2014rfv}, with the uncertainty on their fluxes of the order 1\%, will be able to constrain the $\nu_\mu$ normalisation and $\nu_e\nu_\mu$ triangle far beyond what is currently obtainable. However, no one experiment can probe all scales and complementarity is vital to definitively make a statement about unitarity from new low-energy physics, especially as there is little means to directly measure the $\nu_\tau$ sector. Improvement in $\nu_\tau$ appearance requires new experiments with both an intense, well known beam of high enough energy $\nu_\mu$ or $\nu_e$ to kinematically produce charged taus, as well as a detector technology capable of efficiently identifying them to a degree necessary for precision high statistics measurements, both of which are extremely difficult tasks. Perhaps crucially for $\nu_\tau$ measurements, Hyper-Kamiokande \cite{Abe:2011ts} will be incredibly sensitive to atmospherically averaged steriles, $\geq0.1 \text{ eV}^2$, and will significantly improve the current bounds on $\tnorm$ in this regime, to approximately $1-\tnorm \leq 0.07$ at the 99\% CL \cite{hktalk}, which would bring it closer inline with the other sectors. \\

In this paper we have emphasised the fact that current experimental bounds on unitarity within the $3\nu$ paradigm allows for considerable violation, and without the unitarity assumption, the precision on the individual $U_\text{PMNS}$ elements can vary significantly (up to $104\%$ in the case of $|U_{\tau 3}|$). However, we find no evidence for non-unitarity.  The prospects of directly measuring all the 12 unitarity constraints with high precision are poor, and even when one allows for additional model-dependant sterile searches we can only constrain the amount of non-unitarity to be $\lesssim$ 0.2 - 0.4, for four out of six of the row and columns normalisations,  with the $\nu_\mu$ and $\nu_e$ normalisation deviations from unity constrained to be $\leq$ 0.07, all at the  $3\sigma$ CL, see Fig. \ref{fig:norm}. Similarly, five out of six of the unitarity triangles are only constrained to be $\lesssim$ 0.1 - 0.2, with opening of the remaining $\nu_e \nu_\mu$ triangle being constrained to be $\leq$ 0.03, again at the $3\sigma$ CL, see Fig. \ref{fig:triangle}. One must be careful when assessing the current experimental regime with the addition of new physics we are currently insensitive to, as without the assumption of unitarity there is much room for new effects, especially in the $\nu_\tau$ sector where currently significant information comes from the unitarity assumption and not direct measurements. \\

{{\bf Acknowledgements}: We thank Belen Gavela and Silvia Pascoli for fruitful discussions. MRL would like to thank the Theoretical Physics department at Fermilab for hosting him during much of this work. SP thanks the Kavli Institute for Theoretical Physics in UC Santa Barbara for their hospitalities, where part of this work was done. The authors acknowledge partial support from the European Union FP7 ITN INVISIBLES (Marie Curie Actions, PITN-GA-2011-289442). This research was supported in part by the National Science Foundation under Grant No. NSF PHY11-25915. Fermilab is operated by the Fermi Research Alliance under contract no. DE-AC02-07CH11359 with the U.S. Department of Energy. }

\bibliographystyle{apsrev4-1}
\bibliography{nonunit_New}{}

\end{document}